\documentclass[twocolumn]{aastex631}

\usepackage{savesym}
\savesymbol{tablenum}
\restoresymbol{SIX}{tablenum}
\usepackage{graphicx}
\usepackage{txfonts}
\usepackage{hyperref}

\newcommand{\Ha}{H{$\alpha$}}
\newcommand{\kms}{\rm km\,s$^{-1}$}
\def \kms {{\rm km\,s$^{-1}$}}

\def \msun {{\rm M$_\odot$}}

\def \cmcb {{\rm cm$^{-3}$}}

\newcommand{\JWST}{\rm {\it JWST}}
\newcommand{\HST}{\rm {\it HST}}
\newcommand{\Astrosat}{\rm {\it Astrosat}}

\newcommand{\ch}[1]{\textcolor{black}{ #1}}

\newcommand{\namea}{\rm The Phantom Void}
\newcommand{\nameb}{\rm The Precursor Phantom Void}

\received{October, 2022}
\revised{-, 2021}
\accepted{\today}

\shorttitle{Anatomy of bubbles in NGC628}
\shortauthors{Barnes et al.}

\begin{document}

\title{PHANGS-JWST First Results: Multi-wavelength view of feedback-driven bubbles (The Phantom Voids) across NGC~628}

\author[0000-0003-0410-4504]{Ashley.~T.~Barnes}
\affiliation{Argelander-Institut f\"{u}r Astronomie, Universit\"{a}t Bonn, Auf dem H\"{u}gel 71, 53121, Bonn, Germany}
\affiliation{European Southern Observatory, Karl-Schwarzschild-Stra{\ss}e 2, 85748 Garching, Germany}
\email{ashleybarnes.astro@gmail.com}

\author[0000-0002-7365-5791]{Elizabeth~J.~Watkins}
\affiliation{Astronomisches Rechen-Institut, Zentrum f\"{u}r Astronomie der Universit\"{a}t Heidelberg, M\"{o}nchhofstra\ss e 12-14, 69120 Heidelberg, Germany}


\author[0000-0002-6118-4048]{Sharon E. Meidt}
\affiliation{Sterrenkundig Observatorium, Universiteit Gent, Krijgslaan 281 S9, B-9000 Gent, Belgium}

\author[0000-0001-6551-3091]{Kathryn~Kreckel}
\affiliation{Astronomisches Rechen-Institut, Zentrum f\"{u}r Astronomie der Universit\"{a}t Heidelberg, M\"{o}nchhofstra\ss e 12-14, 69120 Heidelberg, Germany}

\author[0000-0001-6113-6241]{Mattia C. Sormani}
\affiliation{Universit\"{a}t Heidelberg, Zentrum f\"{u}r Astronomie, Institut f\"{u}r Theoretische Astrophysik, Albert-Ueberle-Stra{\ss}e 2, D-69120 Heidelberg, Germany}

\author[0000-0002-9483-7164]{Robin G. Tre{\ss}}
\affiliation{Institute of Physics, Laboratory for galaxy evolution and spectral modelling, EPFL, Observatoire de Sauverny, Chemin Pegais 51, 1290 Versoix, Switzerland.}

\author[0000-0001-6708-1317]{Simon C.~O.\ Glover}
\affiliation{Universit\"{a}t Heidelberg, Zentrum f\"{u}r Astronomie, Institut f\"{u}r Theoretische Astrophysik, Albert-Ueberle-Stra{\ss}e 2, D-69120 Heidelberg, Germany}

\author[0000-0003-0166-9745]{Frank Bigiel}
\affiliation{Argelander-Institut f\"ur Astronomie, Universit\"at Bonn, Auf dem H\"ugel 71, 53121 Bonn, Germany}

\author[0000-0003-0085-4623]{Rupali Chandar}
\affiliation{Ritter Astrophysical Research Center, University of Toledo, Toledo, OH 43606, USA}

\author[0000-0002-6155-7166]{Eric Emsellem}
\affiliation{European Southern Observatory, Karl-Schwarzschild-Stra{\ss}e 2, 85748 Garching, Germany}
\affiliation{Univ Lyon, Univ Lyon1, ENS de Lyon, CNRS, Centre de Recherche Astrophysique de Lyon UMR5574, F-69230 Saint-Genis-Laval France}

\author[0000-0002-2278-9407]{Janice C. Lee}
\affiliation{Gemini Observatory/NSF’s NOIRLab, 950 N. Cherry Avenue, Tucson, AZ, 85719, USA}

\author[0000-0002-2545-1700]{Adam~K.~Leroy}
\affiliation{Department of Astronomy, The Ohio State University, 140 West 18th Avenue, Columbus, Ohio 43210, USA}
\affiliation{Center for Cosmology and Astroparticle Physics, 191 West Woodruff Avenue, Columbus, OH 43210, USA}

\author[0000-0002-4378-8534]{Karin M. Sandstrom}
\affiliation{Department of Physics, University of California, San Diego, 9500 Gilman Drive, San Diego, CA 92093, USA}

\author[0000-0002-3933-7677]{Eva Schinnerer}
\affiliation{Max-Planck-Institut f\"ur Astronomie, K\"onigstuhl 17, D-69117 Heidelberg, Germany}

\author[0000-0002-5204-2259]{Erik Rosolowsky}
\affiliation{Department of Physics, University of Alberta, Edmonton, Alberta, T6G 2E1, Canada}

\author[0000-0002-2545-5752]{Francesco Belfiore}
\affiliation{INAF — Arcetri Astrophysical Observatory, Largo E. Fermi 5, I-50125, Florence, Italy}

\author[0000-0003-4218-3944]{Guillermo A. Blanc}
\affiliation{The Observatories of the Carnegie Institution for Science, 813 Santa Barbara St., Pasadena, CA, USA}
\affiliation{Departamento de Astronom\'ia, Universidad de Chile, Camino del Observatorio 1515, Las Condes, Santiago, Chile}

\author[0000-0003-0946-6176]{Médéric~Boquien}
\affiliation{Centro de Astronomía (CITEVA), Universidad de Antofagasta, Avenida Angamos 601, Antofagasta, Chile}

\author[0000-0002-8760-6157]{Jakob den Brok}
\affiliation{Argelander-Institut f\"{u}r Astronomie, Universit\"{a}t Bonn, Auf dem H\"{u}gel 71, 53121, Bonn, Germany}

\author[0000-0001-5301-1326]{Yixian Cao}
\affiliation{Max-Planck-Institut f\"ur Extraterrestrische Physik (MPE), Giessenbachstr. 1, D-85748 Garching, Germany}

\author[0000-0002-5635-5180]{M\'elanie Chevance}
\affiliation{Universit\"{a}t Heidelberg, Zentrum f\"{u}r Astronomie, Institut f\"{u}r Theoretische Astrophysik, Albert-Ueberle-Stra{\ss}e 2, D-69120 Heidelberg, Germany}
\affiliation{Cosmic Origins Of Life (COOL) Research DAO, coolresearch.io}

\author[0000-0002-5782-9093]{Daniel~A.~Dale}
\affiliation{Department of Physics and Astronomy, University of Wyoming, Laramie, WY 82071, USA}

\author[0000-0002-4755-118X]{Oleg V. Egorov}
\affiliation{Astronomisches Rechen-Institut, Zentrum f\"{u}r Astronomie der Universit\"{a}t Heidelberg, M\"{o}nchhofstra\ss e 12-14, 69120 Heidelberg, Germany}

\author[0000-0002-1185-2810]{Cosima Eibensteiner} 
\affiliation{Argelander-Institut für Astronomie, Universität Bonn, Auf dem Hügel 71, 53121 Bonn, Germany}

\author[0000-0002-3247-5321]{Kathryn Grasha}
\affiliation{Research School of Astronomy and Astrophysics, Australian National University, Canberra, ACT 2611, Australia}   
\affiliation{ARC center of Excellence for All Sky Astrophysics in 3 Dimensions (ASTRO 3D), Australia}  

\author[0000-0002-9768-0246]{Brent Groves}
\affiliation{International center for Radio Astronomy Research, University of Western Australia, 7 Fairway, Crawley, 6009 WA, Australia}

\author[0000-0002-8806-6308]{Hamid Hassani}
\affiliation{Department of Physics, University of Alberta, Edmonton, Alberta, T6G 2E1, Canada}

\author[0000-0001-9656-7682]{Jonathan~D.~Henshaw}
\affiliation{Astrophysics Research Institute, Liverpool John Moores University, 146 Brownlow Hill, Liverpool L3 5RF, UK}
\affiliation{Max-Planck-Institut f\"ur Astronomie, K\"onigstuhl 17, D-69117 Heidelberg, Germany}

\author[0000-0002-4232-0200]{Sarah Jeffreson}
\affiliation{Center for Astrophysics, Harvard \& Smithsonian}

\author[0000-0002-9165-8080]{Mar\'ia J. Jim\'enez-Donaire}
\affiliation{Observatorio Astron\'{o}mico Nacional (IGN), C/Alfonso XII, 3, E-28014 Madrid, Spain}

\author[0000-0002-9642-7193]{Benjamin W. Keller}
\affiliation{Department of Physics and Materials Science, University of Memphis, 3720 Alumni Avenue, Memphis, TN 38152, USA}

\author[0000-0002-0560-3172]{Ralf S.\ Klessen}
\affiliation{Universit\"{a}t Heidelberg, Zentrum f\"{u}r Astronomie, Institut f\"{u}r Theoretische Astrophysik, Albert-Ueberle-Stra{\ss}e 2, D-69120 Heidelberg, Germany}
\affiliation{Universit\"{a}t Heidelberg, Interdisziplin\"{a}res Zentrum f\"{u}r Wissenschaftliches Rechnen, Im Neuenheimer Feld 205, D-69120 Heidelberg, Germany}

\author[0000-0001-9605-780X]{Eric W. Koch}
\affiliation{Center for Astrophysics | Harvard \& Smithsonian, 60 Garden St., 02138 Cambridge, MA, USA}

\author[0000-0002-8804-0212]{J.~M.~Diederik~Kruijssen}
\affiliation{Cosmic Origins Of Life (COOL) Research DAO, coolresearch.io}

\author[0000-0003-3917-6460]{Kirsten L. Larson}
\affiliation{AURA for the European Space Agency (ESA), Space Telescope Science Institute, 3700 San Martin Drive, Baltimore, MD 21218, USA}

\author[0000-0002-4825-9367]{Jing~Li}
\affiliation{Astronomisches Rechen-Institut, Zentrum f\"{u}r Astronomie der Universit\"{a}t Heidelberg, M\"{o}nchhofstra\ss e 12-14, 69120 Heidelberg, Germany}

\author[0000-0001-9773-7479]{Daizhong Liu}
\affiliation{Max-Planck-Institut f\"ur Extraterrestrische Physik (MPE), Giessenbachstr. 1, D-85748 Garching, Germany}

\author[0000-0002-1790-3148]{Laura A. Lopez}
\affiliation{Department of Astronomy, The Ohio State University, 140 West 18th Avenue, Columbus, Ohio 43210, USA}
\affiliation{Center for Cosmology and Astroparticle Physics, 191 West Woodruff Avenue, Columbus, OH 43210, USA}
\affiliation{Flatiron Institute, Center for Computational Astrophysics, NY 10010, USA}

\author[0000-0001-7089-7325]{Eric J.\,Murphy}
\affiliation{National Radio Astronomy Observatory, 520 Edgemont Road, Charlottesville, VA 22903, USA}

\author[0000-0001-9793-6400]{Lukas Neumann}
\affiliation{Argelander-Institut f\"{u}r Astronomie, Universit\"{a}t Bonn, Auf dem H\"{u}gel 71, 53121, Bonn, Germany}

\author[0000-0003-3061-6546]{Jérôme Pety}
\affiliation{IRAM, 300 rue de la Piscine, 38400 Saint Martin d'H\`eres, France}
\affiliation{LERMA, Observatoire de Paris, PSL Research University, CNRS, Sorbonne Universit\'es, 75014 Paris}

\author[0000-0001-5965-3530]{Francesca Pinna}
\affiliation{Max-Planck-Institut f\"ur Astronomie, K\"onigstuhl 17, D-69117 Heidelberg, Germany}

\author[0000-0002-0472-1011]{Miguel~Querejeta}
\affiliation{Observatorio Astron\'{o}mico Nacional (IGN), C/Alfonso XII, 3, E-28014 Madrid, Spain}

\author[0000-0001-5073-2267]{Florent~Renaud}
\affiliation{Department of Astronomy and Theoretical Physics, Lund Observatory, Box 43, SE-221 00 Lund, Sweden} 

\author[0000-0002-2501-9328]{Toshiki Saito}
\affiliation{National Astronomical Observatory of Japan, 2-21-1 Osawa, Mitaka, Tokyo, 181-8588, Japan}

\author[0000-0002-4781-7291]{Sumit K. Sarbadhicary}
\affiliation{Department of Astronomy, The Ohio State University, 140 West 18th Avenue, Columbus, Ohio 43210, USA}
\affiliation{Center for Cosmology and Astroparticle Physics, 191 West Woodruff Avenue, Columbus, OH 43210, USA}

\author[0000-0002-5783-145X]{Amy Sardone}
\affiliation{Department of Astronomy, The Ohio State University, 140 West 18th Avenue, Columbus, OH 43210, USA}

\author[0000-0002-0820-1814]{Rowan J. Smith}
\affiliation{Jodrell Bank center for Astrophysics, Department of Physics and Astronomy, University of Manchester, Oxford Road, Manchester M13 9PL, UK}

\author[0000-0002-0820-1814]{Sophia K. Stuber}
\affiliation{Max-Planck-Institut f\"ur Astronomie, K\"onigstuhl 17, D-69117 Heidelberg, Germany}

\author[0000-0003-0378-4667]{Jiayi~Sun}
\affiliation{Department of Physics and Astronomy, McMaster University, 1280 Main Street West, Hamilton, ON L8S 4M1, Canada}
\affiliation{Canadian Institute for Theoretical Astrophysics (CITA), University of Toronto, 60 St George Street, Toronto, ON M5S 3H8, Canada}

\author[0000-0002-8528-7340]{David A. Thilker}
\affiliation{Department of Physics and Astronomy, The Johns Hopkins University, Baltimore, MD 21218, USA}

\author[0000-0003-1242-505X]{Antonio Usero}
\affiliation{Observatorio Astron\'{o}mico Nacional (IGN), C/Alfonso XII, 3, E-28014 Madrid, Spain}

\author[0000-0002-3784-7032]{Bradley C. Whitmore}
\affiliation{Space Telescope Science Institute, 3700 San Martin Drive, Baltimore, MD, USA}

\author[0000-0002-0012-2142]{Thomas G. Williams}
\affiliation{Sub-department of Astrophysics, Department of Physics, University of Oxford, Keble Road, Oxford OX1 3RH, UK}
\affiliation{Max-Planck-Institut f\"ur Astronomie, K\"onigstuhl 17, D-69117 Heidelberg, Germany}

\suppressAffiliations

\begin{abstract} 
We present a high-resolution view of bubbles within The Phantom Galaxy (NGC~628); a nearby ($\sim$10\,Mpc), star-forming ($\sim$2\,\msun\,yr$^{-1}$), face-on ($i$\,$\sim$\,9$^{\circ}$) grand-design spiral galaxy. 
With new data obtained as part of the PHANGS-\JWST\ treasury program, we perform a detailed case-study of two regions of interest, one of which contains the largest and most prominent bubble in the galaxy  (\namea; over 1\,kpc in diameter), and the other being a smaller region that may be the precursor to such a large bubble (\nameb).
When comparing to matched resolution \Ha\ observations from the Hubble Space Telescope (\HST), we see that the ionized gas is brightest in the shells of both bubbles, and is coincident with the youngest ($\sim$1\,Myr) and most massive ($\sim$10$^{5}$\,\msun) stellar associations.
We also find an older generation ($\sim$20\,Myr) of stellar associations is present within the bubble of \namea. 
From our kinematic analysis of the HI, H$_2$ (CO) and HII gas across \namea, we infer a high expansion speed of around 15 to 50\,\kms.
The large size and high expansion speed of \namea\ suggest that the driving mechanism is sustained stellar feedback due to multiple mechanisms, where early feedback first cleared a bubble (as we observe now in \nameb), and since then SNe have been exploding within the cavity, and have accelerated the shell.
Finally, comparison to simulations shows a striking resemblance to our \JWST\ observations, and suggests that such large-scale stellar feedback-driven bubbles should be common within other galaxies.
\end{abstract}


\section{Introduction} 
\label{sec_int}

\begin{figure*}
    \centering
        \includegraphics[width=\textwidth]{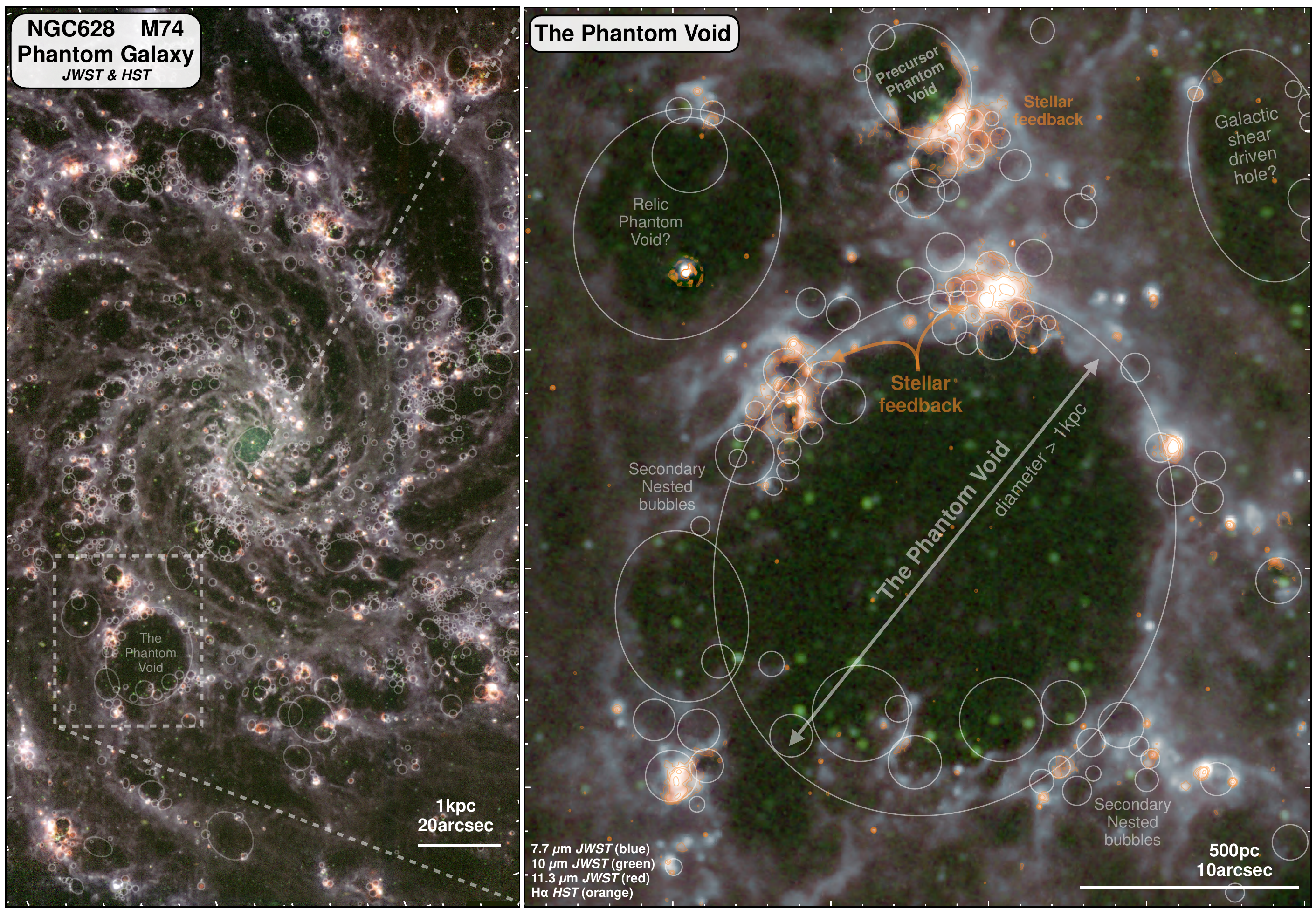}
    \caption{{\bf The prominent bubble structures across The Phantom Galaxy (Messier 74 or NGC~628).} In all panels, we show an image produced from the 770W (blue), 1000W (green), and 1130W (red) band filters from the \JWST\ (Lee et al.\ in prep), and overlaid in orange is the continuum subtracted \HST-\Ha. The faded circles and ellipses show the positions of the bubbles from \cite{WATKINS_PHANGSJWST}.}
    \label{fig2}
\end{figure*}
 
High-mass stars (${>}8$\,\msun) are fundamental in driving the evolution of galaxies, due to the large amounts of energy and momentum (i.e.\ stellar feedback) that they inject into the interstellar medium (ISM) during their short lifetimes \citep[e.g.][]{krumholz_2014}.
At early times (${<}5$\,Myr), feedback in the pre-supernova (pre-SN) stages of high-mass stars (i.e.\ within HII regions) plays a critical role disrupting molecular clouds and forming bubbles within the ISM \citep[e.g.][]{dale_2012, dale_2013, Raskutti2016, Gatto2017, Rahner2017, Rahner2019, Kim2018, Kim_JG2021, Kannan2020, Jeffreson2021, Grasha2018, Grasha2019, Chevance2022b_review, McLeod2021,Barrera-Ballesteros2021a,Barrera-Ballesteros2021b,Barnes2021b, Hannon2019, Hannon2022,Barnes2022}.
At later times, these bubbles can merge \citep{Clarke2002, Simpson2012, Krause2015}, and feedback from supernovae (SNe) can further act to drive expansion, forming so-called (super-)bubble structures with scales of 10s to 1000s of pc over timescales of 10s of Myr \citep[e.g.][]{McKee1977, MacLow1988, oey_superbubble_1997, Weisz2009a, Weisz2009b, Keller2014,Keller2015,Keller2016,kruijssen19a, Chevance2020, Chevance2022a, Kim2021, Kim2022, nath_size_2020, bagetakos_fine-scale_2011, orr_bursting_2022, zucker_star_2022}, even driving fountains out of the galactic plane (e.g. \citealp{veilleux_2005, Fraternali2017}). 

With the advent of the \JWST, we can now, for the first time, achieve the resolution, sensitivity and coverage for an unprecedented view of bubble populations within galaxies beyond the Milky Way and the Local Group.
Specifically, the wavelength coverage of MIRI is perfectly suited to this task, as it is sensitive to several polycyclic aromatic hydrocarbon (PAH) emission features (e.g. at 7.7\,\micron; see \citealp{SMITH07DUST,draine_2007,LI20DUST}). 
Emission from PAHs is particularly useful in tracing the shells of feedback-driven bubbles (e.g. \citealp{Pineda2022}), due to (a) the increased gas densities found in swept-up shells (PAHs are generally well mixed with the gas and illuminated by the average ISRF such that they trace the gas column very sensitively; e.g. \citealt{REGAN06DUST,LEROY13SFGAS,CHOWN21SFGAS,GAO22MIRCO,LEROY1_PHANGSJWST}), (b) the high number of ionizing photons emitted by the OB association powering the bubbles, leading to PAHs being destroyed in the photoionized interiors of the bubbles (e.g. \citealt{GALLIANO18REVIEW, EGOROV_PHANGSJWST, CHASTENET1_PHANGSJWST, CHASTENET2_PHANGSJWST}) and (c) the low optical depth from the shell interior to the edge, which will allow FUV photons to easily heat the small dust grains (e.g. \citealp{draine_2007, draine_2011, Hensley2021}).
Together, these cause the edges of bubbles to appear with high contrast against their interior PAH emission (e.g. \citealp{Churchwell2006,Watson2008}).

In a companion paper in this issue, \cite{WATKINS_PHANGSJWST} used a combination of \JWST\ and \HST\ observations to study the bubble population across the nearby (9.84$\pm$0.63\,Mpc; \citealp{Anand2021a, Anand2021b}), star-forming (1.8\,$\pm$\,0.5\,\msun\,yr$^{-1}$; ), \ch{face-on ($i$\,$\sim$\,9$^{\circ}$; \citealp{Lang2020} and also see \citealp{Blanc2013}), massive ($M_*$=10$^{10.3}$\msun, $M_\mathrm{HI}$=10$^{9.7}$\msun, $M_\mathrm{H2}$=10$^{9.4}$\msun; \citealp{Walter2008, Querejeta2015, Leroy2019, Leroy2021a_survey}),} spiral galaxy Messier 74 (also known as NGC\,628 and The Phantom Galaxy). 
\ch{These authors manually identify bubbles using a combination of 7.7\,\micron\ \JWST-MIRI observations \citep{LEE_PHANGSJWST}, {\it B}-band (438\,nm) PHANGS–\HST\ observations \citep{Lee2021}, and VLT-MUSE \Ha\ observations \citep{Emsellem2022}.
For each bubble, \citet{WATKINS_PHANGSJWST} fit circular or elliptical apertures to the bubble boundaries (i.e. the shells) seen within this multi-wavelength dataset, and in doing so identify $\sim$1700 bubbles with radii ranging between 6 and 500\,pc (e.g. left panel of Fig.\,\ref{fig2}).}
These structures are clearly pervasive across the galaxy, \ch{form a complex nested structure (where smaller bubbles are preferentially located at the edges of large bubbles)}, and are among the most striking features in the initial \JWST\ images (see \citealp{WATKINS_PHANGSJWST} for a in-depth discussion of the nesting of these structures and their size distributions).

In the \cite{WATKINS_PHANGSJWST} catalog, one hole stands out due to its size (over $\sim$\,1\,kpc in diameter), circular shape, and strong contrast with respect to its environment (e.g. right panel of Fig.\,\ref{fig2}). 
We refer to this impressive feature as \namea, which is the focus of this letter (Tab.\,\ref{table:1} summarizes the main properties of \namea\ that have been determined within this work). 
To help understand this structure, we also identify a nearby more compact, but still very well-defined bubble, which we call \nameb\ and analyze in parallel (also see Tab.\,\ref{table:1}).
To conduct our analysis, we combine all the datasets available taken as part of the PHANGS\footnote{\url{http://www.phangs.org}}--JWST survey \citep{LEE_PHANGSJWST} and existing PHANGS multi-wavelength observations to assemble a complete panchromatic, multi-phase picture of the gas, stars, and dust in these bubbles. 

\begin{table}
\caption{{\bf Properties of The Phantom Void and The Precursor Phantom Void.} Tabulated are the properties of the bubble (i.e. the ellipsoid central cavity), and shell (i.e. the ellipsoid annullus around the cavity) of each source (shown in Fig.\,\ref{fig3}). We present the central position, the semi-major and semi-minor axis length and the position angle of the ellipse used to define the outer boundary of the bubbles and shells, and, also, the mean radius of these ellipses in units of parsec. We also present the total molecular (\S\,\ref{sec_expansion_mol}) and atomic (\S\,\ref{sec_expansion_atom}) hydrogen masses and mass surface densities, and total stellar mass in young stellar associations (derived from the association catalog of \citealp{Deger2020} and Larson et al.\ subm.; \S\,\ref{sec_stellarbubs}). Lastly, we show an estimate of the expansion velocity (\S\,\ref{sec_expandbubs}).}              
\label{table:1}      
\centering                                      
\begin{tabular}{l c c c c}          
\hline\hline                        
Property & \multicolumn{2}{c}{\namea} & \multicolumn{2}{c}{Precursor} \\
 & Bubble & Shell & Bubble & Shell \\
\hline                  

RA [deg] 	 	 & 24.1866 	 & 24.1864 	 & 24.1863 	 & 24.1863 \\
Dec [deg] 	 	 & 15.7719 	 & 15.7719 	 & 15.7784 	 & 15.7782 \\
$r_\mathrm{major}$ [arcsec] 	 	 & 10.8 	 & 17.6 	 & 1.7 	 & 4.9 \\
$r_\mathrm{minor}$ [arcsec] 	 	 & 4.5 	 & 16.0 	 & 1.3 	 & 4.3 \\
$r_\mathrm{pa}$ [deg] 	 	 & 126 	 & 126 	 & 30 	 & 30 \\
$r_\mathrm{mean}$ [pc] 	 	 & 364 	 & 801 	 & 69 	 & 219 \\
M$_\mathrm{H2}$ [M$_\odot$/10$^5$] 	 	 & 12.8 	 & 379.9 	 & 3.1 	 & 38.0 \\
M$_\mathrm{HI}$ [M$_\odot$/10$^5$] 	 	 & 7.0 	 & 56.4 	 & 0.2 	 & 3.5 \\
M$_\mathrm{*}$ [M$_\odot$/10$^5$] 	 	 & 1.8 	 & 7.8 	 & 1.4 	 & 1.4 \\
$\Sigma_\mathrm{H2}$ [M$_\odot$\,pc$^{-2}$] 	 	 & 3.7 	 & 22.8 	 & 20.6 	 & 28.1 \\
$\Sigma_\mathrm{HI}$ [M$_\odot$\,pc$^{-2}$] 	 	 & 2.1 	 & 3.4 	 & 2.4 	 & 2.6 \\

$v_{\exp}$ [\kms] & \multicolumn{2}{c}{$\sim$\,20} & \multicolumn{2}{c}{$\sim$\,6} \\
\hline                                             
\end{tabular}
\end{table}

\section{Observations} 
\label{sec_obs}

\subsection{PHANGS-JWST observations} 
\label{sec_obs_jwst}

The PHANGS (Physics at High Angular resolution in Nearby GalaxieS)-\JWST\ observations were taken as part of the Cycle~1 treasury project ID 02107 \citep{LEE_PHANGSJWST}, which targets 19 nearby, star-forming galaxies with NIRCam (F200W, F300M, F335M and F360M) and MIRI (F770W, F1000W, F1130W and F2100W) imaging.
The observations targeting NGC~628 cover the main star-forming disk (containing 50\% of the total star formation of the galaxy), which is matched to coverage from Hubble \citep{Lee2021}, VLT-MUSE \citep{Emsellem2022}, and ALMA \citep{Leroy2021a_survey}. 
We primarily make use of the F770W filter observations in this work, which have a point spread function (PSF) full width at half maximum (FWHM) of $\sim$\,0.25\arcsec\ ($\sim$12\,pc at the galaxy distance). 
A detailed description of the complete data reduction is presented in \citet{LEE_PHANGSJWST}.

\subsection{HST observations} 
\label{sec_obs_hst}


We make use of NUV–{\it U–B–V–I} band \HST\ observations taken from the LEGUS survey \citep{Calzetti2015}, and reduced using the PHANGS-HST survey pipeline (see \citealp{Lee2021}). 
The PSF of these observations have a FWHM of $\sim$\,0.1-0.2\arcsec\ ($\sim$5-10\,pc). In addition, we use the narrow-band F658N map from the \HST, to produce a higher resolution ($\sim$\,0.1\arcsec) \Ha\ emission map (Proposal 10402).
To do so, the F658N map is continuum subtracted using an image formed from a combination of the F814W and F550M maps, appropriately scaled using their AB zero-points (see \citealp{Hannon2022} for methods).\footnote{We are not correcting the narrow-band flux for the contribution of the [NII] emission lines. This has no impact on the results presented in this letter as we are primarily interested in the morphology of the emission rather than its absolute brightness.}

In this letter, we also include the properties (e.g.\ ages and masses; \S\,\ref{sec_stellarbubs}) taken from the stellar association catalog (\citealp{Deger2020, Lee2021}, Larson et al.\ subm). 
The properties in this catalog were determined from SED modeling of the \HST\ broadband filters with CIGALE \citep{Boquien2019} based on the \citet{Bruzual2003} single stellar populations, while also including ionized gas emission (lines and continuum) and dust attenuation. 
We make use of the {\it B} band selected catalog with an association scale of 32\,pc (Larson et al.\ subm). 
For this work, using a catalog selected from a different band, or association over a different scale, makes no significant difference to our results.

\subsection{Ancillary observations} 

We also make use of continuum subtracted emission line maps (\Ha\, [OI]6300, [SII]6716) and line kinematics based on VLT/MUSE observations from the PHANGS-MUSE survey (see \citealp{Emsellem2022} for a complete discussion of the processing and reduction of these observations). These provide a higher sensitivity, yet lower resolution ($\sim$\,1\arcsec), view of the ionized gas compared to the \HST\ \Ha\ map -- better suited to identifying the diffuse emission within bubbles.
In addition, we use CO\,(2-1) observations from the PHANGS-ALMA survey (see \citealp{Leroy2021a_survey} for a complete description of the survey and \citealp{Leroy2021b_pipeline} for the processing and reduction of the data), a NUV emission map from Astrosat (Hassani et al.\ in prep), a 8\,\micron\ map from {\it Spitzer} \citep{Kennicutt2003,Dale2009}, and natural weighted HI observations from the VLA (THINGS survey; \citealp{Walter2008}).
We conduct our analysis by using a common astrometric grid, retaining the native resolution of our multi-wavelength datasets to preserve information (as opposed to smoothing to a common resolution).

\section{Blowing Bubbles} 
\label{sec_bubsample}

\begin{figure*}[!t]
    \centering
    \includegraphics[width=\textwidth]{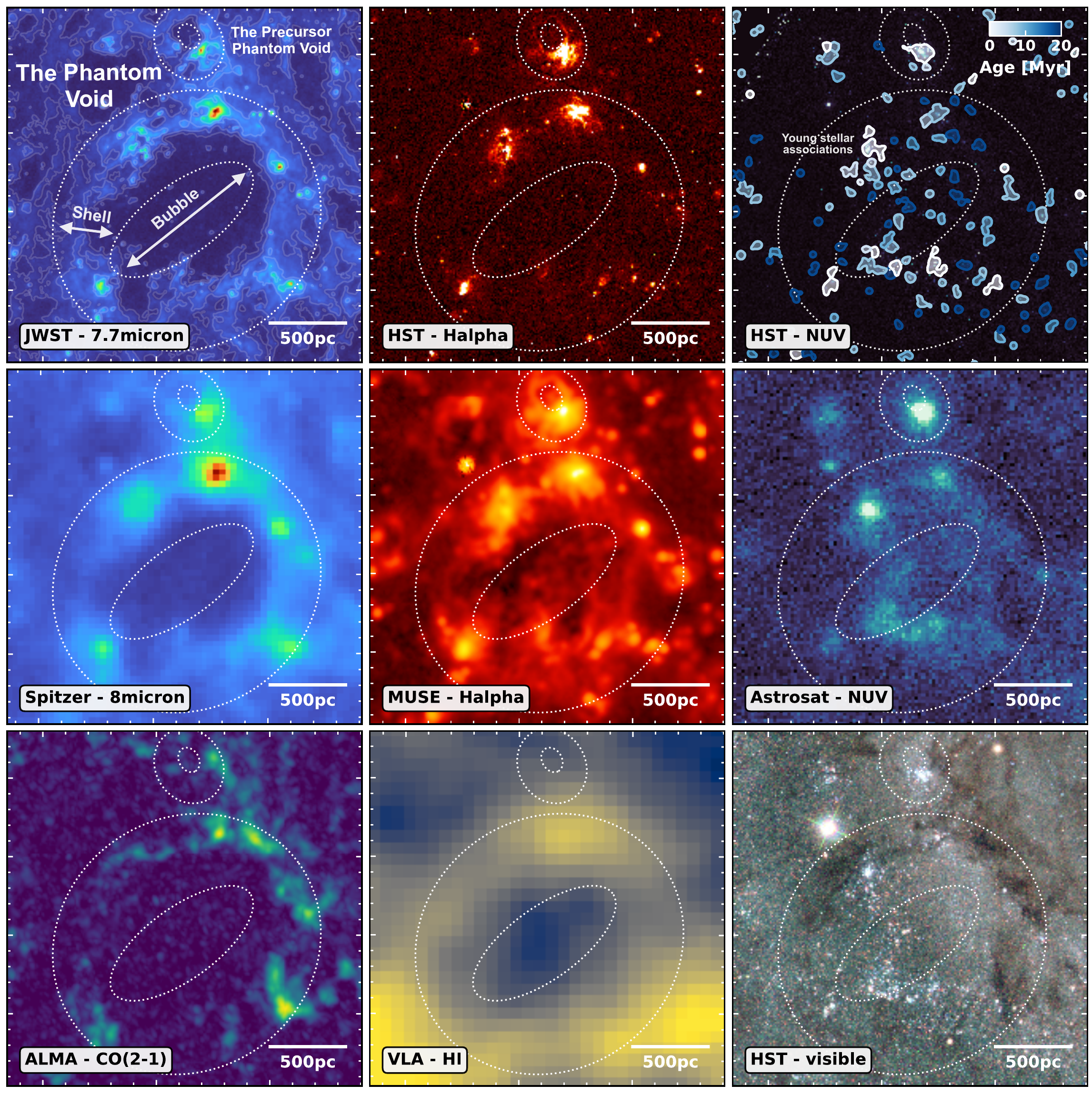}
    \caption{{\bf Comparison of a feedback-driven bubble, and its host stellar population for the Phantom Void and the Precursor Phantom Void (also shown in Fig.\,\ref{fig2}).} From left to right in upper panels, we show the \JWST\ 7.7\,\micron, \HST\ \Ha\ and the near-UV (F275W) filters. Overlaid as colored contours on the upper right panel is the age of the stellar associations (at a scale of 32\,pc; see \citealp{Deger2020}). From left to right in center panels are the {\it Spitzer 8\,\micron} \citep{Dale2009}, MUSE \Ha\ \citep{Emsellem2022}, and Astrosat NUV (Hassani et al.\ in prep) observations. In the bottom panels are the ALMA CO \citep{leroy_phangs-alma_2021}, VLA HI \citep{Walter2008}, and HST broadband \citep{Lee2021} observations.
    The white dashed line on all panels denotes the bubble and shell of each region (see Fig.~\ref{fig5}). We find that young and high-mass stellar associations ($<$\,20\,Myr; $>$10$^{5}$\,\msun) exist within the bubble (particularly towards the boundaries), highlighting these as good candidates for driving the bubble expansion.}
    \label{fig3}
\end{figure*}

\subsection{Detailed look at feedback bubbles} 
\label{sec_identbubs}

\begin{figure}[!t]
    \centering
    \includegraphics[width=\columnwidth]{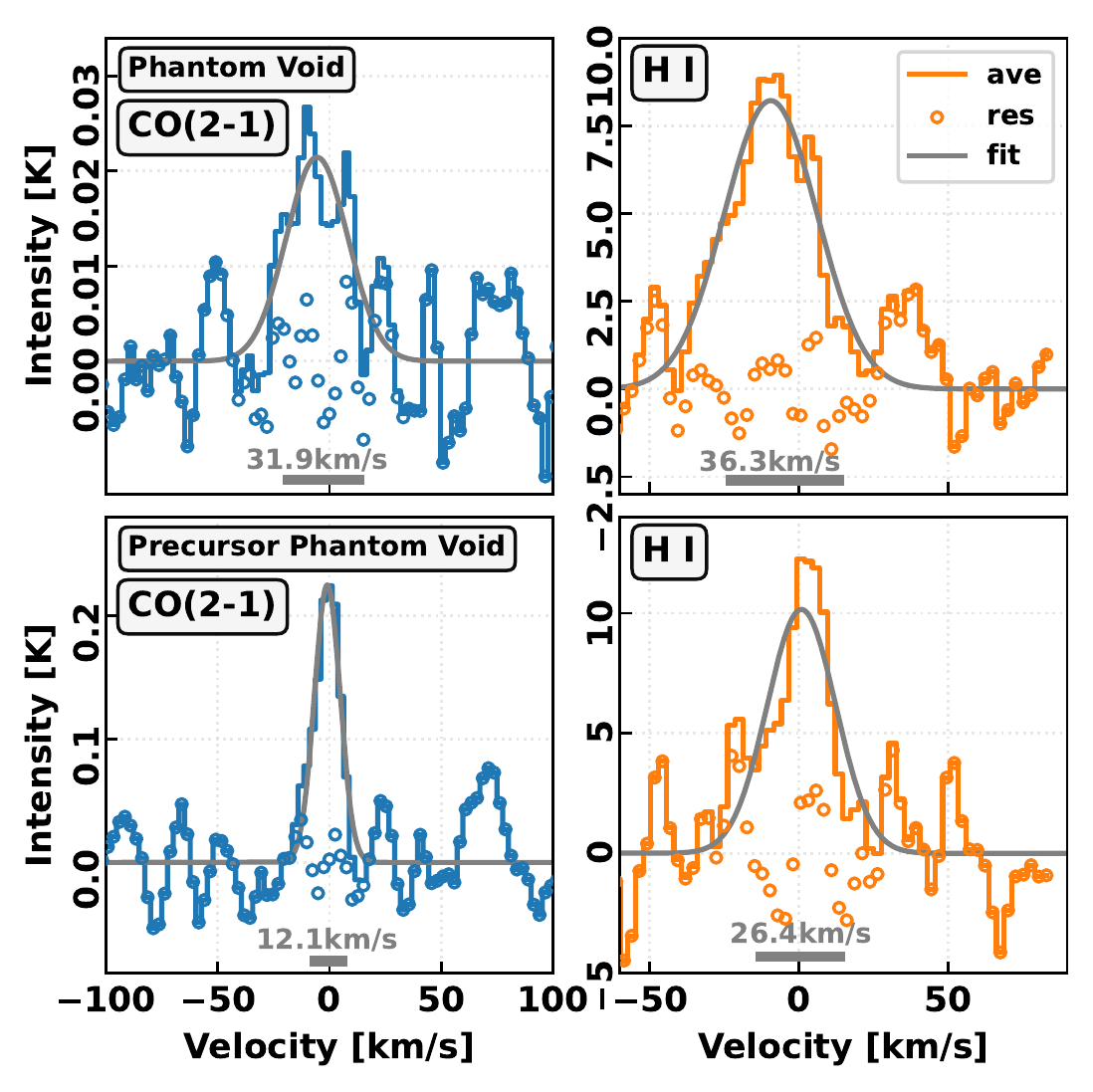}
    \caption{{\bf Spectra taken inside of the Phantom Void bubble (see Fig.~\ref{fig3}).} In the left and right panels, we present the average CO\,(2-1) spectrum obtained from the cube that has had the local velocity field subtracted (including rotational and non-rotational components), and the average HI spectrum, respectively. Gaussian fits to the data are overlaid as gray curves. Residuals are plotted as open points. Below each curve, we show a bar and text showing the measured FWHM (corrected for the instrument spectral resolution).}
    \label{figspec}
\end{figure}

Fig.\,\ref{fig3} present a detailed, multi-wavelength view of \namea\ and \nameb\ regions.
In \namea, we see that the main bubble is relatively devoid of any \JWST\ 7.7\,\micron\ emission (values of $<$1\,MJy\,sr\,pix$^{-1}$) or \HST-\Ha\ emission (values of $<$10$^{-17}$\,erg\,s$^{-1}$\,cm$^{-1}$\,arcsec$^{-2}$) towards the central cavity, and emission is discretely distributed around the bubble shell. 
We note that we use this nomenclature throughout this letter, where bubble refers to the inside cavity, while the shell refers to the perimeter with some thickness (as labeled on Fig.\,\ref{fig3}).
Comparing to the broadband near-UV \HST\ and Astrosat observations, we find that many of these emission peaks in the shell are connected to compact stellar emission sources (discussed further in \S\,\ref{sec_stellarbubs}).
When comparing to the high-sensitivity, but lower resolution, PHANGS-MUSE observations, we see that there is a low level component to the diffuse \Ha\ emission inside of the bubble.
This emission could be connected to the more diffuse emission seen in broadband \HST\ and Astrosat observations, and is consistent with ionization by an older generation of stars (with respect to those associated with compact \Ha\ emission in the shell). 

In contrast, \nameb\ appears to be associated with a single star-forming complex (seen in \Ha\ and near-UV emission) located at the edge of the shell. 
Moreover, there is clear evidence from the \HST\ observations that the ionizing radiation is propagating through the bubble and illuminating its far edge, which appears bright in \Ha\ emission.
Low level \Ha\ is also seen in the higher sensitivity MUSE observations within the bubble cavity. 
Additionally, it appears that there are many more compact and less evacuated secondary bubbles towards the south of the region, which are associated with bright extended \Ha\ emission.    

Together, the structures seen in \namea\ and \nameb\ suggest that the bubbles we are seeing in the \JWST\ imaging are driven by multiple generations of high-mass stars, the youngest of which are located at the bubble edges (see \S\,\ref{sec_starformaiton} for a discussion of e.g. trigger star formation). 
These stars are then injecting their ionizing radiation into the bubble, which must have a relatively low density of hydrogen such that the radiation can propagate through the bubble and cause the diffuse \Ha\ emission at the dusty shell boundary. 
Specifically, for a characteristic ionizing photon energy of 20~eV, we expect the optical depth of a 100~pc bubble with mean atomic hydrogen density $n_{\rm H}$ to be approximately $\tau \sim 500 \, (n_{\rm H} / {\rm cm^{-3}})$, implying that $n_{\rm H}$ cannot be larger than $\sim 0.01$\,\cmcb.\footnote{\ch{Here we make the assumption that all the gas is atomic. However it could be possible that some of the gas is ionized, which would lower the atomic gas optical depth and allow for a higher density.}}
That said, \namea\ is even larger than this, so could have an even lower density. 
Indeed, \namea\ is so large that it is even visible as a hole within comparatively low resolution HI observations ($\sim$\,500\,pc beam; \citealp{Walter2008}), confirming a low atomic hydrogen density within its evacuated center (see Fig.~\ref{fig3}). 
Given that the estimated HI disk scale height of NGC\,628 is $\sim$\,800\,pc \citep{Dutta2008}, roughly the size of the \namea, it has likely burst out of the galaxy. 
Bubbles on this scale are one mechanism that could be responsible for shaping galactic-scale chimneys \citep{Heiles1984} and establishing gas recycling via galactic fountains (e.g. \citealp{Fraternali2017}). 
An interesting avenue for the future would be to investigate if this cavity is filled with outflowing hot X-ray gas (e.g. with Chandra). 

\subsection{Bubble expansion} 
\label{sec_expandbubs}

We aim to measure the expansion speed of \ch{the bubbles} with a number of independent methods using our multi-wavelength datasets. 
These are summarized in the following subsections. 

\subsubsection{Molecular gas} 
\label{sec_expansion_mol}

We make two estimates of the expansion velocity from the CO\,(2-1) data \citep{leroy_phangs-alma_2021}. 
First, we make an estimate using the line-width measured inside of the bubble, which probes the motion out of the plane (i.e. a difference between faint emission at the front and back side of the bubble). 
Second, we make a estimate using the velocity of the emission in the shell of the bubble, which probes the motion in the direction of the galactic plane. 

In Fig\,\ref{figspec}, we show the average CO\,(2-1) spectrum taken from inside of both bubbles, which show a weak, yet significant, signal.
Here, we use a data cube that has had the local velocity field subtracted, including projected circular and non-circular motions, and, hence, the signal is centered on $\sim$\,0\,\kms.
We fit the emission with a Gaussian function, and measure a Full Width at Half Maximum (FWHM) of 31.9\,\kms\ for \namea, and 12.1\,\kms\ for \nameb.
We make the assumption that this line-width is the result of the line-of-sight component of the approaching and receding side of the bubble shell.\footnote{We note, however, that for large bubbles of the order of galactic scale heights, such as the Phantom Void, the bubble expansion deforms into a prolate spheroid perpendicular to the galaxy disc. This can effect the measured expansion velocity with respect to the expansion in the plane of the galaxy (e.g. see \citealp{Baumgartner2013}).}
Assuming spherical expansion, then $v_{\exp}=\mathrm{FWHM/2}\,\sim\,16$\,\kms\ for \namea\ and $\sim\,6$\,\kms\ for \nameb, where the expansion velocity here is the difference in velocity between the receding and approaching side of the bubble out of the plane. 

\begin{figure}[!t]
    \centering
    \includegraphics[width=1\columnwidth]{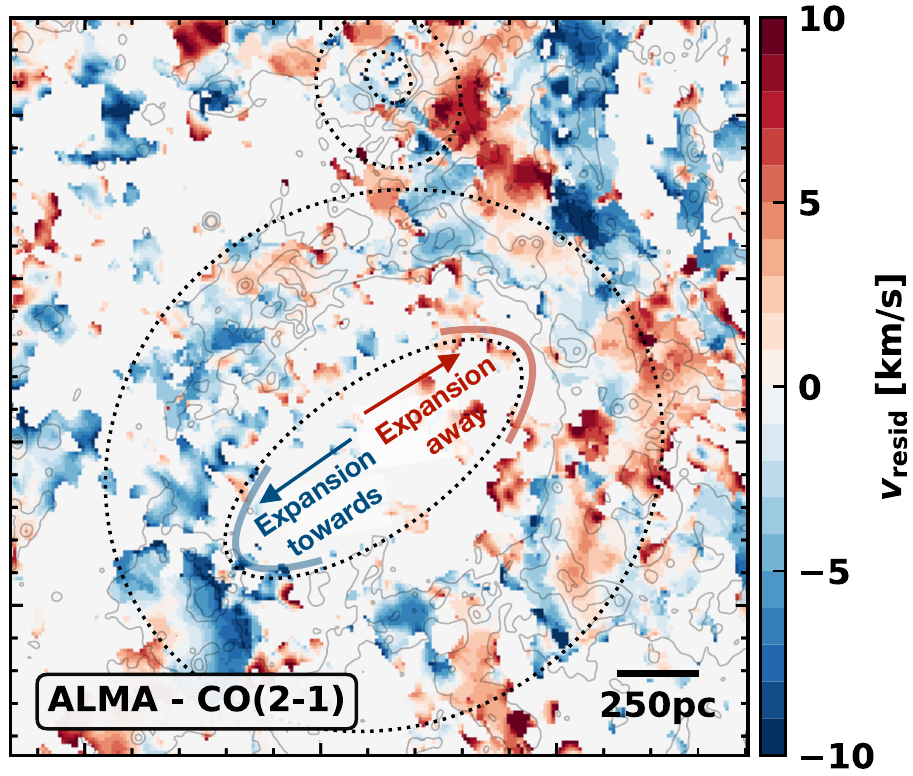}
    \caption{{\bf Residual CO velocity towards the voids.} Here we show the local, non-rotational flows at the location of the phantom void, constructed from the intensity weighed centroid velocity map after subtracting the projected rotational velocity (measured by \citealt{Lang2020}) from each pixel. Overlaid as black contours on each panel is the 7.7\,\micron\ emission (see Fig.\,\ref{fig3}).}
    \label{figmoms_shuff}
\end{figure}

In Fig.\,\ref{figmoms_shuff}, we show the molecular gas velocity field that highlights local, non-circular flows in the galactic plane at the location of \nameb.
Focusing on the centroid velocity of the emission around the bubble, we find that residual velocities are negative on the left side of the shell and positive on the right side of the shell. 
Given the orientation, rotation direction and inclination of NGC\,628, this would correspond to expansion outward from the center of the bubble (i.e. the left part of the shell is moving towards a larger galactocentric radius and the right part of the shell is moving towards a smaller galactocentric radius).
Specifically, from Fig\,\ref{figmoms_shuff}, we measure $v_\mathrm{resid}$ on the left part of the bubble is $\sim$\,$-$5\,\kms, whereas on the right part of the bubble it is +3\,\kms, with a few positions at +5\,\kms.
This would give $|v_\mathrm{resid}|\sim$3$-$5\,\kms, and hence we estimate that for \namea\  $v_\mathrm{exp}=|v_\mathrm{resid}|/\mathrm{sin}(i)$\,=\,20$-$35\,\kms (where $i=8.9^{\circ}$; \citealp{Lang2020}; also see \citealp{Blanc2013}). 

Given the simplistic assumptions on the geometry of the system, we consider that these estimates of the expansion speed perpendicular and parallel to the plane of the galaxy are consistent for the \namea.
Hence the expansion velocity of the molecular gas is taken to be in the range $v_{\exp}\sim15 -35$\,\kms. 
\ch{As \nameb\ is located within the spiral arm, and due to its smaller size relative to the \namea, there is more confusion in separating the bubble from the surrounding gas within the residual velocity map. 
Hence, the in-plane expansion speed of \nameb\ cannot be estimated, and we take the expansion speed estimate from the out-of-plane method: $v_{\exp}\sim6$\,\kms.}

\begin{figure*}[!t]
    \centering
    \includegraphics[width=\textwidth]{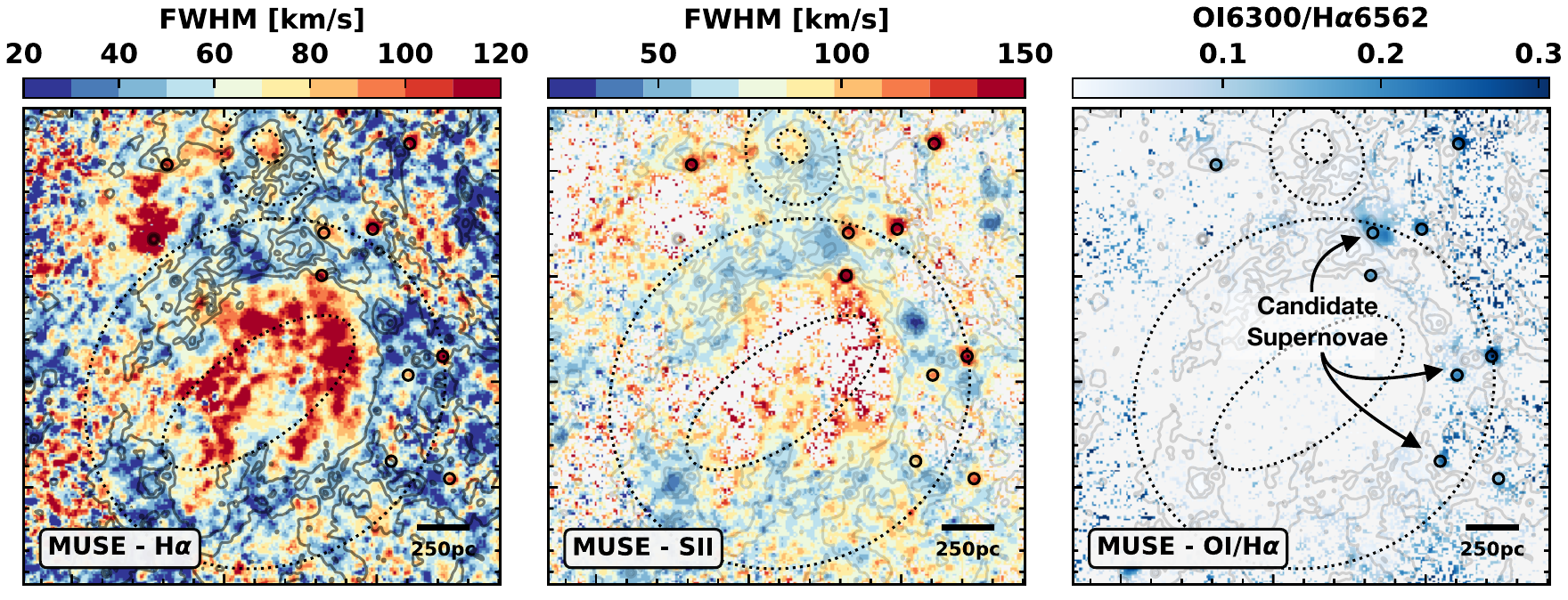}
    \caption{{\bf Evidence of stellar feedback in the ionized gas tracers towards the voids.} From left to right, we show the emission-weighted velocity FWHM maps from the \Ha\ and the [SII]6716 lines, and the emission line ratio [OI]6300/\Ha. Overlaid as black contours on each panel is the 7.7\,\micron\ emission (see Fig.\,\ref{fig3}). We see elevated line widths distributed throughout the center of the bubble (albeit at low signal-to-noise), which we attribute to turbulence and out-of-plane expansion. Moreover, as highlighted on each panel, we see elevated line widths ($>$100\,\kms) and line ratios towards the shell, which could be indicative of young supernova remnants (Li et al. in prep).} 
    \label{figmoms}
\end{figure*}


\subsubsection{Atomic gas} 
\label{sec_expansion_atom}

To constrain the expansion speed in the neutral gas, we use the VLA HI (natural weighted) data cube (Fig.\,\ref{fig3}). 
As with the CO\,(2-1), we measure the averaged spectrum for the center of \namea\ bubble (Fig.\,\ref{figspec} right panel), and again fit the emission with a Gaussian function. 
We measure a FWHM of 36.3\,\kms, which is very similar to the estimate obtained from the CO (32.4\,\kms).
This gives a $v_{\exp}$\,=\,$\mathrm{FWHM/2}$\,$\sim$\,18\,\kms.

For completeness, we also estimate the atomic gas expansion speed of \nameb\ using the same method. 
We measure FWHM=26.4\,\kms, giving $v_{\exp}$\,$\sim$\,13\,\kms. 
This estimate, should, however, be taken with caution, as the beam size of the VLA HI is larger than the size of \nameb\ bubble, and, hence, could include a significant contribution from gas kinematics outside of the bubble.

\subsubsection{Ionized gas} 
\label{sec_expansion_ion}

We use the MUSE \Ha\ emission line kinematics to constrain the expansion speed in the ionized gas (Fig\,\ref{figmoms}, left panels). 
Across the same region used to measure the molecular and atomic expansion speeds, we find that the average FWHM for \namea\ is 88.4\,\kms\ ($v_{\exp}\,\sim\,45$\,\kms).\footnote{The velocity dispersion shown Fig.\,\ref{figmoms} has been corrected for the instrumental velocity dispersion \citep{Emsellem2022}, and converted to the FWHM (factor of $\sqrt{8ln(2)}$).}
\ch{Similarly, we find a FWHM for \nameb\ of 85.1\,\kms\ ($v_{\exp}\,\sim\,45$\,\kms).}
These values are significantly higher than what is measured in the neutral gas ($\sim$\,6-20\,\kms), even when accounting for the larger thermal contribution to the FWHM of the warmer ionized medium (e.g $\sim$\,20\,\kms\ at 10$^4$\,K), which could highlight the fact that the ionized medium is more directly connected to the source of feedback. 

Interestingly, we find systematically elevated line-widths in the interior of both bubbles relative to their shells, and even measure FWHM values of more than 100\,\kms in \namea\ (though at these very low S/N$<$10 note there is a bias towards overestimating line-widths). 
Such high values of the line-width could be indicative of supernova feedback contributing to increased turbulence within the center of the bubble (Egorov et al. in prep). 
Near the shell, we directly identify supernova remnants via their broadened [SII]6716 line emission and increased [OI]/\Ha\ line ratios (Fig.\,\ref{figmoms}; Li et al, in prep).  
The presence of SN explosions within the shell could be contributing to continued driving of the bubble expansion (\S\,\ref{sec_stellarbubs}).

\subsubsection{Galactic dynamical constraints}
\label{sec_expandbubsdyn}

\begin{figure}[!t]
    \centering
    \includegraphics[width=\columnwidth]{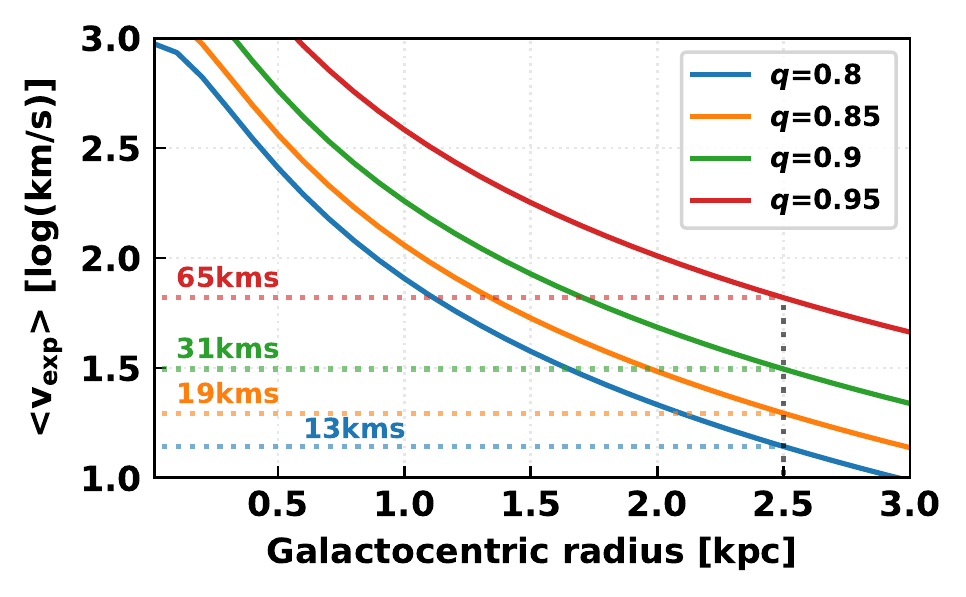}
    \caption{{\bf The expansion velocity required for Phantom Void-sized shells to retain a high degree of circularity as they expand in the presence of differential rotation at different locations in the inner disk of NGC\,628.} Curves show the $v_\mathrm{exp}$ required for different observed axis ratios $q$=0.8 to 0.95 (in steps of 0.05). Here a bubble radius $R_{\rm b}$=500 pc is adopted and NGC 628's rotation curve is modeled with a two-parameter fit to the measured rotation curve \citep{Lang2020}. Highlighted are the results corresponding to the approximate galactocentric radius of \namea.}
    \label{figshear}
\end{figure}

In the presence of differential rotation or `rotation curve shear', expanding structures will become distorted over time \citep{Palous1990}.  
Here we use the ellipticity $\epsilon$ of shells of a given radius to place rough constraints on their expansion velocity.
Assuming an isotropic bubble expansion velocity, $v_\mathrm{exp}$, and bubble radius, $R_\mathrm{b}=\int_0^t v_{\rm exp}\,dt$ (in the absence of shear), after bubble age $t$, the axis ratio ($q=1-\epsilon$) of shearing bubbles can be written
\begin{equation}
q=\left(1+\frac{dV_\mathrm{c}}{dR}t_\mathrm{b}\right)^{-1}=\left(1+\frac{dV_\mathrm{c}}{dR}\frac{\langle R_\mathrm{b}\rangle}{\langle v_\mathrm{exp}\rangle}\right)^{-1} \label{eq:shearingbubbles}
\end{equation}
in terms of the rotation curve derivative $dV_\mathrm{c}/dR$, the average (time-weighted) bubble radius 
\begin{equation}
\langle R_\mathrm{b}\rangle=\frac{\int_0^t R_\mathrm{b} dt}{t},
\end{equation}
the average (time-weighted) bubble expansion velocity 
\begin{equation}
\langle v_{\rm exp}\rangle=\frac{R_b}{t}=\frac{\int_0^t v_{\rm exp} dt}{t},
\end{equation}
(as would be estimated from the present day size and age)
and effective bubble age as estimated by $t_\mathrm{b}$=$\langle R_\mathrm{b}\rangle$/$\langle v_\mathrm{exp}\rangle$=$\langle R_\mathrm{b}\rangle$/$R_\mathrm{b} t$. 

Using Eq. (\ref{eq:shearingbubbles}), the $v_\mathrm{exp}$ required to generate bubbles with a given set of average sizes and ellipticities can be determined, provided that the rotation curve is known, that is,  
\begin{equation}
\langle v_{\rm exp}\rangle=\frac{dV_\mathrm{c}}{dR} \langle R_\mathrm{b}\rangle\left(\frac{1}{q}-1\right)^{-1}.\label{eqexpansionwshear}
\end{equation}

\noindent Below we use this expression to obtain a basic estimate of $v_{\rm exp}$ under the assumption of negligible time evolution in the bubble expansion rate, such that the current bubble radius is a good approximation of $<$$R_b$$>$.  Other assumptions for the time evolution of $v_{\rm exp}$ will yield different results, but this should be useful for illustration purposes.\footnote{For expansion that declines over time, $\langle v_{\rm exp}\rangle$ will overestimate the present day expansion velocity.  For example, in the event of an exponentially decreasing $v_{\rm exp}$ with time i.e. $v_{\rm exp}(t)=v_0e^{-t/t_0}$ with characteristic time $t_0$, then when the evolution is fast and $t_0<<t$, the stalled radius $R_b$ is well approximated by $\langle R_b\rangle\sim v_0t_0 \sim R_b$ and the expansion velocity measured from eq. (\ref{eqexpansionwshear})  $\langle v_{\rm exp}\rangle=v_0t_0/t=R_b/t$ exceeds the present day $v_{\rm exp}$ but is representative of the expansion near its fastest.}

The Phantom void, with a radius of $R_\mathrm{b}\sim$500 pc, sits at $R_\mathrm{gal}\sim$2.5 kpc and has a low ellipticity corresponding to $q\sim0.85$.  
Fig. \ref{figshear} shows the expansion velocities that would be required for an expanding structure with these properties to retain such a high degree of circularity.  
Here we have adopted an analytical two-parameter model for this galaxy's rotation curve to estimate the rotation curve shear at all locations \citep{Lang2020}.  
To produce a structure like the Phantom void, we estimate expansion velocities $v_\mathrm{exp}\sim$\,20\,\kms are required.

\subsubsection{Mass and energy of the shell} 

We use the CO\,(2-1) ALMA observations to estimate that there is 3.8\,$\times$\,10$^7$\,\msun\ of molecular gas within the shell of \namea\ (1.3\,$\times$\,10$^6$\,\msun\ inside the bubble). 
Here we use the same shell mask shown in Fig.\,\ref{fig3}, adopt a CO\,(2-1)/CO\,(1-0) ratio of 0.61 based on direct observations \citep{denBrok2021}, and a $\alpha_\mathrm{CO(1-0)}=3.93$\,\msun\,pc$^{-2}$\,(\kms)$^{-1}$ as predicted by empirical scaling relations \citep{sun20}.
In addition, we estimate that there is 5.6\,$\times$\,10$^6$\,\msun\ of atomic gas within the shell of \namea\ (7.0\,$\times$\,10$^5$\,\msun\ inside the bubble), when using the VLA data and the conversion presented in \citet{Bigiel2010}.

When taking a representative expansion speed determined in the previous section of $v_\mathrm{exp}\sim$\,20\,\kms, we calculate that the energy required to drive a shell with this combined atomic and molecule mass is 1.7\,$\times$\,10$^{53}$\,erg.
We find the total ($\mathrm{HI}+\mathrm{H}_2$) mass inside the shell of \nameb\ is about ten times less than the shell (4.2\,$\times$\,10$^6$\,\msun), which -- when assuming the estimated expansion speed of $v_\mathrm{exp}\sim$\,6\,\kms -- implies a kinetic energy of $\sim$\,10$^{51}$\,erg.

\ch{A plausible energy source for these bubbles is stellar feedback (as discussed further in the following section). 
For example, a typical value for the amount of kinetic energy released per supernova is $\sim$10$^{51}$\,erg (see e.g. \citealp{Bethe1990,draine_2011}), and, hence, approximately 100 SNe would be required to power \namea.
On the other hand, \nameb\ could be powered by a single SN.
However, it is worth noting that this number could vary significantly depending on, for example, the contribution from pre-supernova feedback (e.g. in the form of winds, which over the lifetime of a high-mass star have been suggested to provide an energy contribution similar to a SN; e.g. \citealp{Chevance2022b_review}), and coupling efficiency of these feedback mechanisms with the expanding bubble (e.g. values of the amount of kinetic energy retained into the surrounding medium, and not radiated away, ranges between a few to a few ten percent; e.g. \citealp{Thornton1998,Tamburro2009,Sharma2014,Kim2015a,Martizzi2015,Gentry2017}).
We estimate the mass of a stellar association required to produce these number of SNe assuming $f_{* \rightarrow \mathrm{SN}}/\left<m\right> \sim 0.01$, or that one SN is produced per 100\msun\ of stars that fully populate an initial mass function (see \citealp{Tamburro2009}). 
Hence, we estimate that \namea\ should contain $>$\,10$^{4}$\,\msun\ of stars with ages less than a few 10 Myr (i.e. the time scale for the stars to explode as SNe after they are formed), and \nameb\ $>$\,10$^{2}$\,\msun. As we show in the following section, these criteria are fulfilled.}


\subsection{Stellar populations driving bubbles} 
\label{sec_stellarbubs}

\begin{figure}
    \centering
    \includegraphics[width=\columnwidth]{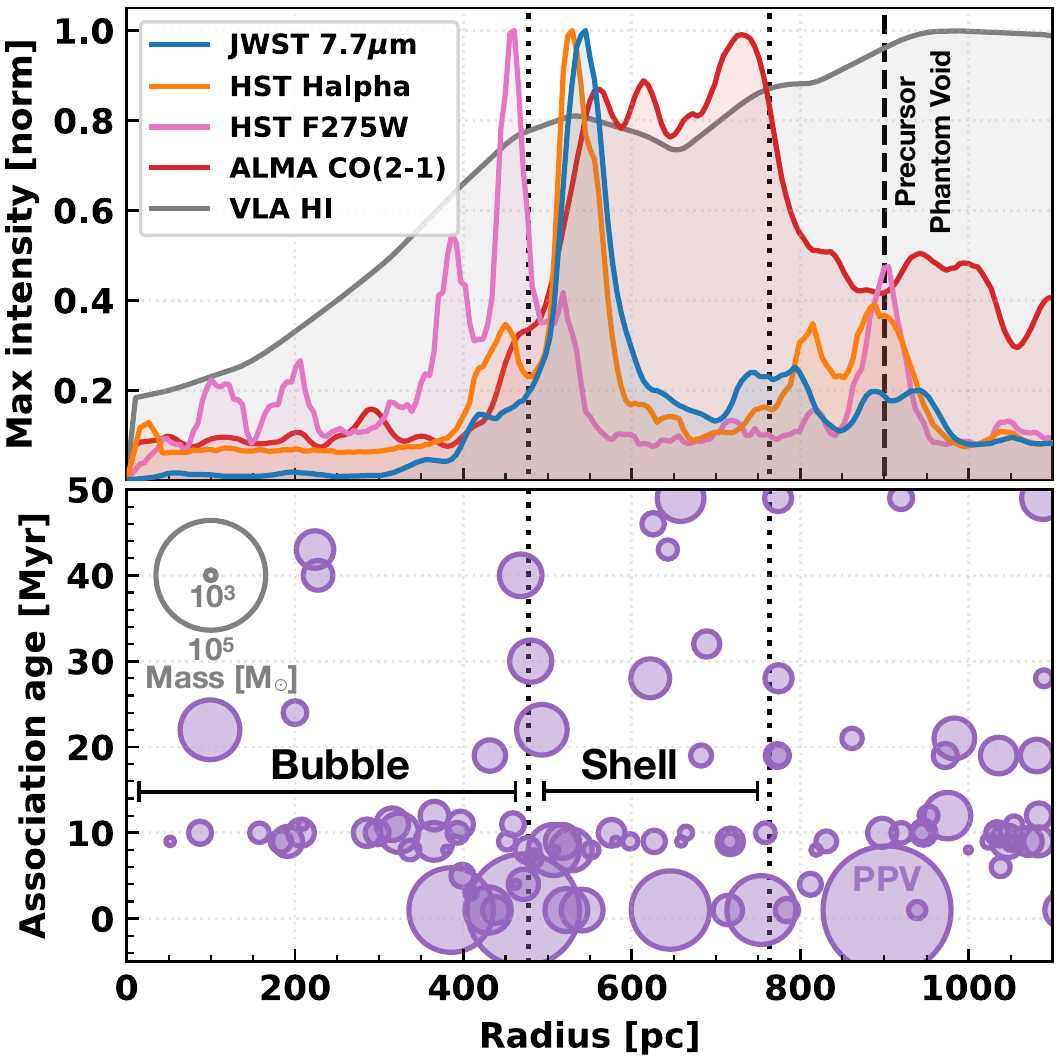}
    \caption{{\bf Distribution of intensity and stellar population in the Phantom Void (see Fig.~\ref{fig3}).} The upper panel shows the intensity distribution of the \JWST\ 7.7\,\micron,  continuum subtracted \Ha\ emission, \HST\ NUV, ALMA CO, and VLA HI emission, as measured radially from the center of the bubble (see Fig.~\ref{fig3}). We show the 99$^{\rm th}$ percentile of the intensity in each radial bin, normalized to the value of the maximum bin (this percentile is chosen instead e.g.\ the mean value to better highlight where the emission distributions peak). The bottom panel shows the radial age distribution of the stellar association catalog. The size of the points indicate the mass of the stellar associations.  In both panels we highlight the position of the \nameb. We find evidence that younger and more higher-mass clusters are preferentially located within the shell, with respect to inside the bubble.}
    \label{fig5}
\end{figure}

Within \namea, we find many compact and bright concentrations of NUV bright (young) stellar point sources within the shell (see \HST\ F275W filter in Fig.\,\ref{fig3}), which spatially correspond to regions of bright \Ha\ and 7.7\micron\ emission. 
In addition, there appears to be an overabundance (with respect to a similar size interarm region adjacent to the bubble) of fainter point sources distributed throughout the bubble cavity, which is also seen as diffuse emission in the Astrosat NUV map. 

Fig.~\ref{fig5} (upper panel) shows the flux distribution of the \JWST\ 7.7\,\micron, \HST\ \Ha, \HST\ NUV (F275W) filter, ALMA CO, and VLA HI emission maps measured radially from the center of \namea.  
We find a good correspondence between the 7.7\,\micron\ and \Ha\ emission, which both peak within the shell radius at $\sim$ 600\,pc (compare to dashed circles in Fig.\,\ref{fig3}). We also see that the \HST\ NUV (F275W) emission is fainter within the bubble.
Interestingly, we find that the peak in the NUV emission is at a smaller ($\sim$\,100\,pc) radius to the 7.7\,\micron\ and \Ha\ emission, hinting that these clusters may have already evacuated their immediate environment of dense atomic gas and ionized gas, respectively. 
We also see that the neutral gas tracers (CO and HI emission) are increased within the shell. 
\nameb, on the other hand, shows a single compact cluster of point sources in the shell (again correspondent with bright \Ha\ and 7.7\,\micron\ emission), and very little point source emission within the bubble.

In Fig.\,\ref{fig3} (upper right panel), we also overlay as colored contours the ages of the stellar associations (see \S\,\ref{sec_obs_hst}; \citealp{Deger2020}, Larson et al.\ subm).  
These indicate that the stellar associations throughout the bubble and shell are generally young ($<$\,20\,Myr). 
We estimate the total stellar mass of these young stellar associations within the \namea\ region to be $10^{6.0}$\,\msun\ (80\% of this stellar mass is in the shell, and 20\% is in the bubble).
These associations could be viable powering sources for these bubbles, either currently or, in the case of those inside the bubble of \namea, they may represent the past generation of stars that created the bubbles (\S\,\ref{sec_starformaiton}).

Interestingly, several of the stellar associations are very young ($<$\,5\,Myr). 
Both in \namea\ and \nameb, these associations reside in the shells (as inferred from the \Ha). 
Indeed, Fig.~\ref{fig5} (lower panel) shows the radial distribution of ages from the association catalog across \namea\, where the size of each point indicates the stellar mass. 
We see evidence that there is a higher number of younger and more massive associations situated within the shell, as opposed to inside of the bubble.
We find that there are 10 associations with ages less than 5\,Myr within the shell of \namea, which have a maximum mass of 10$^{5}$\,\msun\ (total mass of 3.2\,$\times$10$^{5}$\,\,\msun), and, therefore, are massive enough to harbor OB stars (i.e.\ strong sources of stellar feedback).


We consider whether the stellar population ages align with the previous estimates of the bubble expansion speed. 
To test this, we make a calculation of the expansion velocity based on the radius of the bubble ($R \sim$\,500\,pc), and the mean age of the associations within the bubble of $t \sim$\,10\,Myr. 
According to the classical \cite{weaver_1977} model of expanding bubbles, $v_{\rm exp} \simeq 0.6 R/t$ assuming they are in an adiabatic stage of their evolution. 
This implies $v_{\rm exp}$\,=\,30~\kms, which is very close to the values 
estimated from our kinematic analysis (\S\,\ref{sec_expandbubs}).

The large size, high expansion speed, and large energy required to power the \namea\ bubble (\S\,\ref{sec_expandbubs}) suggests that the driving mechanism is stellar feedback, sustained by a combination of multiple mechanisms.
We propose that early feedback first cleared a bubble, as we are now currently witnessing for \nameb\ (and to some extent continue to see in the shell of \namea).
Since then, SNe have been exploding within the cavity, and have accelerated the shell (e.g.\ \citealp{Keller2022}). 
This sustained injection of energy and momentum is what is required reach high (50\,\kms) expansion speeds (i.e. seen in the ionized gas), because the continuous energy input keeps the pressure in the bubble high, whereas having only a single SN would cause the pressure to drop as the bubble expands (e.g.\ \citealp{MacLow1988}).
Finally, the star formation and SNe that are observed in the shell could further contribute to the current and future expansion of the shell.

\section{Star formation at the bubble boundaries} 
\label{sec_starformaiton}

We have found that \namea\ contains a significant amount of ongoing star formation within its shell (as seen in the \Ha\ emission and the young, massive stellar associations), and also that this shell contains a large reservoir of molecular gas (see Tab.\,\ref{table:1}). 
This could then imply several physical scenarios as to why star formation is preferentially located towards the shell, of which we provide an non-exhaustive list of examples below. 
\begin{itemize}
    \item[i)] Gravitationally unbound gas is collected in the shell of the bubble as it expands, which then becomes dense and then star formation is triggered  (i.e.\ gas that would not have otherwise formed stars; e.g. \citealp{Elmegreen1977, Whitworth1994}). 
    \item[ii)] Gravitationally bound gas either from the parent cloud, or surrounding the initial star forming region, is moved by the expansion of the bubble into the shell, which then forms stars (i.e.\ gas that would have otherwise formed stars, but in a different location). 
    \item[iii)] If star formation proceeds sequentially in a cloud/region in an inside-out fashion, then we would naturally get the oldest stars in the middle of the cloud, with younger stars near the edge. 

\end{itemize}

Differentiating between these cause and effect scenarios for the formation of the feedback-driven bubbles and future episodes of star formation is non-trivial, and has been a very long-standing problem in the Galactic star formation community. 
Nonetheless, below we outline methods to test this in future work.

The different cases make different predictions regarding the ages of the various young stellar clusters/associations.
In scenario (i), where star formation is triggered by shell collapse, we would expect all of the young regions in the shell to have ages less than the age of the shell (and hence also less than the age of the central associations). 
However, this need not be the case in scenario (ii). 
If bound clouds are being swept-up, that are already capable of forming stars, then it is likely that some of them will already have stars when they get swept up, whereas others will not, and we would therefore expect to see a broader distribution of ages than in scenario (i).

Another interesting test would be to look at whether there is a population of young stellar clusters/associations just inside the shell. 
This is what one might expect if the expanding shell sweeps away molecular clouds that have already started forming stars, as the shock will act on the gas but not the stars, which should therefore get left behind. 
This would be much harder to bring about in scenario (i), as in this case, we would expect that the radial outward velocity of the stars at their birth should be the same as that of their parent giant molecular clouds (i.e. they should move along with the shell, or even move out faster than the shell in the case where it is significantly decelerating).
Interesting, as seen in Fig.\,\ref{fig5}, we do find that the young stars (seen in the \HST-NUV and in the cluster association catalog) are preferentially located on the inner edge of the shell (offset from the peak in 7.7\,\micron\ emission by 50 to 100\,pc). 


\section{Are Phantom Voids common in galaxies?} 
\label{sec_sims}

\begin{figure}[!t]
    \centering
    \includegraphics[width=\columnwidth]{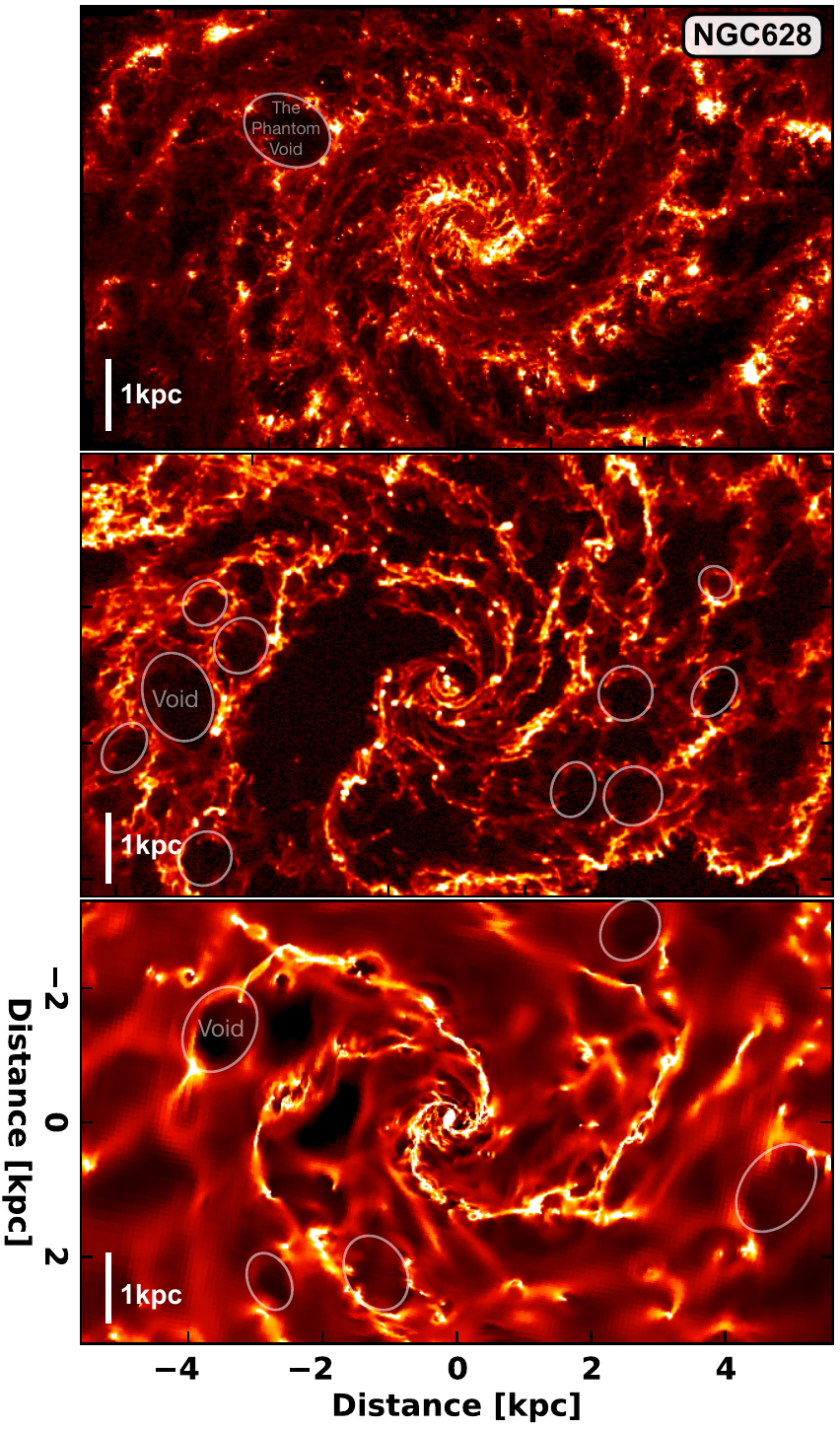}
    \caption{{\bf Phantom Voids in both observations and simulations.} Shown in the upper panel is the \JWST\ 7.7\,\micron\ emission map. The middle panel shows the isolated galaxy simulation from \citet{Tress2020} (the same simulation shown in the bottom panels of their Fig.\ 14). In color-scale is the total column density, which we expect (away from regions of active star formation), to correspond well to the 7.7\,\micron\ emission \citep{SANDSTROM1_PHANGSJWST, LEROY1_PHANGSJWST}. The bottom panel shows the isolated galaxy simulation tuned for NGC\,628 from Emsellem et al.\ (in prep). We have optimized the color-scales such that both maps have a similar qualitative appearance, \ch{and the physical scales of all panels are matched}. We see that large-scale ($\sim$\,1kpc-sized) voids are also common within the simulations, \ch{several of which are highlighted by the faded white ellipses}.}
    \label{fig_sims}
\end{figure}

\namea\ is one of the most striking features within the new \JWST\ data, which was not as pronounced in our existing PHANGS multi-wavelength datasets. 
Only with the resolution and sensitivity afforded by \JWST\ do such coherent, large-scale features become clearly apparent (Fig.\,\ref{fig2}). 
It is then interesting to consider if Phantom Voids are a common feature of other galaxies.

\cite{WATKINS_PHANGSJWST} estimates that there should be $\sim$\,1900 bubbles identifiable in the PHANGS-\JWST\ observations per 1\,\msun\,yr$^{-1}$ of star formation rate (\S\,\ref{sec_identbubs}).
Hence, using their size distribution, we expect to find $\sim$\,1 with a radius of $\sim$500\,pc per 1\,\msun\,yr$^{-1}$ of star formation rate. 
This is more-or-less in line with the observations, given that the star formation rate contained within the mapped region is $\sim$1\,\msun\,yr$^{-1}$. 
Across the remaining galaxies to be mapped as part of the PHANGS-\JWST\ treasury program (total 42\,\msun\,yr$^{-1}$), we could then expect to uncover many Phantom Voids (albeit these may be harder to identify in galaxies with higher inclinations than NGC\,628, because of greater foreground confusion).

We offer a brief comparison between the new JWST observations and theoretical predictions/simulations. Figure\, \ref{fig_sims} compares our 7.7\,\micron\ emission map and snapshots from two sets of galaxy simulations (with matched physical scales). 

The first is the isolated (non-interacting) galaxy simulation from \cite{Tress2020,Tress2021}. This simulation represents a generic isolated galaxy and is not tuned to reproduce the particular properties of NGC\,628. The aim of the simulation was to study molecular clouds and their associated star formation in a galactic environment. The simulation includes a live N-body stellar and dark matter component. The ISM is modeled using a time-dependent chemical network that keeps track of hydrogen and carbon chemistry, a physically motivated model for the formation of new stars using sink particles, and type Ia and type II supernova feedback. The simulation reaches sub-parsec resolution in the dense regions (see Figure 3 in \citealt{Tress2020}) and self-consistently follows the formation of individual molecular clouds from the large-scale flow and their embedded star formation. More details about the simulation can be found in \cite{Tress2020}.

The second simulation is a dedicated hydro-dynamical simulation of an isolated main-sequence star-forming galaxy, using known properties of NGC\,628 to generate initial conditions (stellar and gas mass profiles, velocity profile, viewing angles). The simulation was run using Ramses \citep{Teyssier2002}, an adaptive-mesh refinement code with a maximum sampling of 3.6~pc for the gas cells. It includes live particles for dark matter, old and new stars, atomic and molecular cooling, constant efficiency per free-fall time star formation above a given gas density, with recipes for UV background, feedback from type-Ia and II supernovae, radiative pressure and stellar winds.  

The similarity in morphology between simulations and observations is striking. 
\ch{We show as ellipses in the bottom two panels of Fig.\,\ref{fig_sims} several large-scale ($\sim$\,1kpc-sized) voids, similar to \namea, that have been identified in the simulations.\footnote{We note, however, the identification procedure in the simulations not identical to the observations (which combined multiple filters). In future, we aim to make a more direct comparison based on e.g. using POLARIS to create synthetic observations of the \JWST\ filters as predicted by the simulation.}
In addition, we see many smaller bubbles in the simulations that are similar in size to the \nameb.}
Indeed, simulations have been predicting a bubble-rich ISM morphology for decades \citep[e.g.][]{Wada2008, Grisdale2017}, and HI observations have hinted at a similar morphology (see for example Fig.\ 1 in \citealt{Boomsma2008}). However, HI observations have limited resolution outside of the Milky Way and it is only with the advent of JWST that we can for the first time confirm the predictions of simulations so spectacularly. In simulations which include perturbations from satellites, such as the M51 analog shown in \citet{Pettitt2017}, large-scale voids are also clear in the gas column density. The voids seen in \citet{Pettitt2017} are driven by SN feedback, and also track the edges of the spiral arms, as we see in M74. The strong similarity between observations and simulations suggests that the physical processes included in the simulations (gas self-gravity, supernova feedback, and differential rotation) are at the origin of the observed morphology. Further analysis, that is out of the scope of this paper, will be necessary to quantify in more detail the statistical properties of simulations vs observations (e.g. see \citealp{SANDSTROM1_PHANGSJWST}).

\section{Conclusions} 
\label{sec:conclude}
 

This letter demonstrates  the potential for new PHANGS-\JWST\ observation to be used to identify (morphologically) stellar feedback powered bubbles, and highlights the need for multi-wavelength datasets to study and understand their physical properties. 
Here we provide a detailed case-study of two regions of interest, one of which contains the most prominent bubble in the galaxy (\namea; over 1\,kpc in diameter), and the other being a smaller region that may be the precursor of the larger bubble (\nameb).
We see that the compact \Ha\ typically sits on the edge of the shells, highlighting that the youngest stars are forming at the boundaries of the bubbles -- coincident with the youngest ($\sim$1\,Myr) and most massive ($\sim$10$^{5}$\,\msun) \HST\ stellar associations that are also found towards these regions.
We also find an older generation ($\sim$20\,Myr) of stellar associations that is present within the bubble of \namea. 
From our kinematic analysis of the HI, H$_2$ (CO) and HII gas across \namea, we infer a high expansion speed of around 15 to 50\,\kms.
The large size and high expansion speed of \namea\ suggest that the driving mechanism is sustained stellar feedback, due to multiple mechanisms. 
We propose a scenario where early feedback first cleared a bubble (as we observe now in \nameb), and since then SNe have been exploding within this cavity, and have accelerated the shell (which can also be enhanced by shear).
Finally, comparison to simulations shows a striking resemblance to our \JWST\ observations, and suggest that such massive stellar feedback driven bubbles should be common with other galaxies.

\section*{Acknowledgments}
    
    We would like to thank the referee for their constructive feedback that helped improve the quality of this paper.

    This work was carried out as part of the PHANGS collaboration.
    This work is based on observations made with the NASA/ESA/CSA JWST and NASA/ESA Hubble Space Telescopes. 
    
    Some of the data presented in this paper were obtained from the Mikulski Archive for Space Telescopes (MAST) at the Space Telescope Science Institute, which is operated by the Association of Universities for Research in Astronomy, Inc., under NASA contract NAS 5-03127 for JWST and NASA contract NAS 5-26555 for HST. The JWST observations are associated with program 2107, and those from HST with program 15454. \ch{The specific observations analyzed can be accessed via \dataset[10.17909/9bdf-jn24]{http://dx.doi.org/10.17909/9bdf-jn24} and \dataset[10.17909/t9-r08f-dq31]{https://dx.doi.org/10.17909/t9-r08f-dq31}.}

    This work also makes use of observations collected at the European Southern Observatory under ESO programmes 094.C-0623 (PI: Kreckel), 095.C-0473,  098.C-0484 (PI: Blanc), 1100.B-0651 (PHANGS-MUSE; PI: Schinnerer), as well as 094.B-0321 (MAGNUM; PI: Marconi), 099.B-0242, 0100.B-0116, 098.B-0551 (MAD; PI: Carollo) and 097.B-0640 (TIMER; PI: Gadotti). 
    This publication uses the data from the AstroSat mission and the UVIT instrument of the Indian Space Research Organisation (ISRO), archived at the Indian Space Science Data center (ISSDC). This work is supported by a grant 19ASTROSA2 from the Canadian Space Agency.
    
    This paper makes use of the following ALMA data: ADS/JAO.ALMA\#2012.1.00650.S, ADS/JAO.ALMA\#2017.1.00886.L. ALMA is a partnership of ESO (representing its member states), NSF (USA) and NINS (Japan), together with NRC (Canada), MOST and ASIAA (Taiwan), and KASI (Republic of Korea), in cooperation with the Republic of Chile. The Joint ALMA Observatory is operated by ESO, AUI/NRAO and NAOJ.

    ATB and FB would like to acknowledge funding from the European Research Council (ERC) under the European Union’s Horizon 2020 research and innovation programme (grant agreement No.726384/Empire).
    EJW acknowledges the funding provided by the Deutsche Forschungsgemeinschaft (DFG, German Research Foundation) -- Project-ID 138713538 -- SFB 881 (``The Milky Way System'', subproject P1). 
    JPe acknowledges support by the DAOISM grant ANR-21-CE31-0010 and by the Programme National ``Physique et Chimie du Milieu Interstellaire'' (PCMI) of CNRS/INSU with INC/INP, co-funded by CEA and CNES.
    EWK acknowledges support from the Smithsonian Institution as a Submillimeter Array (SMA) Fellow and the Natural Sciences and Engineering Research Council of Canada.
    JMDK gratefully acknowledges funding from the European Research Council (ERC) under the European Union's Horizon 2020 research and innovation programme via the ERC Starting Grant MUSTANG (grant agreement number 714907). COOL Research DAO is a Decentralized Autonomous Organization supporting research in astrophysics aimed at uncovering our cosmic origins.
    MC gratefully acknowledges funding from the DFG through an Emmy Noether Research Group (grant number CH2137/1-1).
    RSK acknowledges funding from the European Research Council via the ERC Synergy Grant ``ECOGAL'' (project ID 855130), from the Deutsche Forschungsgemeinschaft (DFG) via the Collaborative Research Center ``The Milky Way System''  (SFB 881 -- funding ID 138713538 -- subprojects A1, B1, B2 and B8) and from the Heidelberg Cluster of Excellence (EXC 2181 - 390900948) ``STRUCTURES'', funded by the German Excellence Strategy. RSK also thanks the German Ministry for Economic Affairs and Climate Action for funding in the project ``MAINN'' (funding ID 50OO2206). 
    TGW acknowledges funding from the European Research Council (ERC) under the European Union’s Horizon 2020 research and innovation programme (grant agreement No. 694343). 
    KK, OE gratefully acknowledge funding from the Deutsche Forschungsgemeinschaft (DFG, German Research Foundation) in the form of an Emmy Noether Research Group (grant number KR4598/2-1, PI Kreckel).
    G.A.B. acknowledges the support from ANID Basal project FB210003. SJ is supported by Harvard University through the ITC.
    MB acknowledges support from FONDECYT regular grant 1211000 and by the ANID BASAL project FB210003.
    ER acknowledges the support of the Natural Sciences and Engineering Research Council of Canada (NSERC), funding reference number RGPIN-2022-03499.
    This research was supported by the Excellence Cluster ORIGINS which is funded by the Deutsche Forschungsgemeinschaft (DFG, German Research Foundation) under Germany's Excellence Strategy - EXC-2094-390783311. Some of the simulations in this paper have been carried out on the computing facilities of the Computational Center for Particle and Astrophysics (C2PAP). EE would like to thank Alexey Krukau and Margarita Petkova for their support through C2PAP.
    KG is supported by the Australian Research Council through the Discovery Early Career Researcher Award (DECRA) Fellowship DE220100766 funded by the Australian Government. 
    KG is supported by the Australian Research Council Centre of Excellence for All Sky Astrophysics in 3 Dimensions (ASTRO~3D), through project number CE170100013. 
    MQ acknowledges support from the Spanish grant PID2019-106027GA-C44, funded by MCIN/AEI/10.13039/501100011033.
    FR acknowledges support from the Knut and Alice Wallenberg Foundation.
    CE acknowledges funding from the Deutsche Forschungsgemeinschaft (DFG) Sachbeihilfe, grant number BI1546/3-1
    AKL gratefully acknowledges support by grants 1653300 and 2205628 from the National Science Foundation, by award JWST-GO-02107.009-A, and by a Humboldt Research Award from the Alexander von Humboldt Foundation.
    JS acknowledges support by the Natural Sciences and Engineering Research Council of Canada (NSERC) through a Canadian Institute for Theoretical Astrophysics (CITA) National Fellowship.
    SKS acknowledges financial support from the German Research Foundation (DFG) via Sino-German research grant SCHI 536/11-1.

%

\vspace{5mm}
\facilities{\HST\ (Hubble Space Telescope), \JWST, ALMA (Atacama Large Millimeter/submillimeter Array), VLT-MUSE (Very Large Telescope - Multi Unit Spectroscopic Explorer), VLA (The Karl G. Jansky Very Large Array), {\it Spitzer} space telescope, \Astrosat\ (Astronomy satellite)}


\software{Astropy \citep{AstropyCollaboration2013, AstropyCollaboration2018, AstropyCollaboration2022}, \ch{SAOImageDS9 \citep{Joye2003}, CARTA \citep{Comrie2021}, APLpy \citep{aplpy2012, aplpy2019}}}





\bibliography{main,phangsjwst}{}
\bibliographystyle{aasjournal}


\suppressAffiliationsfalse
\allauthors
\end{document}